\documentclass[]{fairmeta_gr2}

\usepackage{multirow}
\usepackage{color}
\usepackage{colortbl}
\usepackage{mdframed}
\usepackage{caption}
\usepackage{subcaption}
\usepackage{algpseudocode}
\usepackage{algorithm}
\usepackage{amsthm}
\usepackage{amssymb}
\usepackage{graphicx}
\usepackage{enumitem}
\usepackage{xspace}
\usepackage[normalem]{ulem}

\usepackage{booktabs}

\usepackage[most]{tcolorbox}
\usepackage{microtype}   

\newcommand{\approach}{GR2\xspace}

\title{GR2 Technical Report}

\author[*]{Yufei Li}
\author[*]{Zaiwei Zhang}
\author[*]{Mingfu Liang}
\author[*]{Kavosh Asadi}
\author[]{Jay Xu}
\author[]{Jimmy Kim}
\author[]{Chongyang Bai}
\author[]{Jieyi Zhang}
\author[]{Hongye Xie}
\author[]{Prachi Agrawal}
\author[]{Dian Yu}
\author[]{Tianyi Chen}
\author[]{Jean-Pascal Billaud}
\author[]{Garret Buell}
\author[]{YK (Yongkang) Zhu}
\author[]{Sachin Patil}
\author[]{Brooke Bian}
\author[]{Zhou Fang}
\author[]{Kevin Huang}
\author[]{Shiva Sudanagunta}
\author[]{Yuzhen Huang}
\author[]{Emma Lu}
\author[]{Chris O'Brien}
\author[]{Yang Song}
\author[]{Lihong Li}
\author[]{Jacob Tao}
\author[]{Zhicheng Zhu}
\author[]{Chao Li}
\author[]{Gaoxiang Liu}
\author[]{Neil Wu}
\author[]{Zhongyin Hu}
\author[]{Li Han}
\author[]{Loki Chen}
\author[]{Ming Lei}
\author[]{Greg Rehm}
\author[]{Siyuan Song}
\author[]{Tianwei Zhang}
\author[]{Li Li}
\author[]{Ketan Singh}
\author[]{Yavuz Yetim}
\author[]{Ilyas Atishev}
\author[]{Satendra Gera}
\author[]{Ashkan Sadeghi}
\author[]{Rachel Yan}
\author[]{Nikko Mizutani}
\author[]{Shuaiwen Wang}
\author[]{Song Yang}
\author[]{Zhijing Li}
\author[]{Jiang Liu}
\author[]{Mengying Sun}
\author[]{Fei Tian}
\author[]{Xiaohan Wei}
\author[]{Chonglin Sun}
\author[]{Shuo Gu}
\author[]{Parish Aggarwal}
\author[]{Senthil Manickavelu}
\author[]{Kaushik Rangadurai}
\author[]{Zhi Hua}
\author[]{Frank Shyu}
\author[]{Ruchit Sharma}
\author[]{Liyuan Li}
\author[]{Shike Mei}
\author[]{Wenlin Chen}
\author[]{Santanu Kolay}
\author[]{Ben Schulte}
\author[]{Deepak Chandra}
\author[]{Adam (Yang) Song}
\author[]{Sandeep Pandey}
\author[1]{Xi Liu}
\author[2]{Hamed Firooz}
\author[2]{Luke Simon}%

\affiliation[]{Meta AI}

\contribution[*]{Co-first author}
\contribution[1]{Project lead}
\contribution[2]{Direction lead}

\abstract{Industrial recommendation systems serve billions of users through a multi-stage funnel---retrieval, early-stage ranking, and re-ranking---where the final re-ranking step disproportionately shapes user engagement and downstream performance, particularly for carousel and grid display formats.
  Despite growing enthusiasm for Large Language Models (LLMs) in recommendation, three gaps hinder industrial adoption: (1) most efforts target retrieval and ranking, leaving re-ranking---the stage closest to the final user experience---largely underexplored; (2) LLMs are typically deployed zero-shot or via supervised fine-tuning, underutilizing the reasoning capabilities unlocked by reinforcement learning (RL) on verifiable rewards; (3) deployed catalogs index billions of items with non-semantic identifiers that lie outside any base-LLM vocabulary.
  We present \approach (Generative Reasoning Re-Ranker), an end-to-end framework that combines (i)~mid-training on semantic IDs produced by a tokenizer with $\geq$99\% uniqueness, (ii)~reasoning-trace distilled from a stronger teacher via targeted prompting and rejection sampling, and (iii)~RL with verifiable rewards purpose-built for re-ranking.
  To make \approach resource-viable, we further (iv) introduce a context compressor that amortizes training cost, On-Policy Distillation (OPD) as a scalable alternative to SFT---which we find collapses at industrial scale---and reasoning distillation for low-latency serving.
  \approach delivers +18.7\% R@1, +7.1\% R@3, and +9.6\% N@3 over legacy baselines on industrial-scale traffic.
  We further find that reward design is critical in re-ranking: LLMs often hack rewards by preserving the incoming order or exploiting position bias, motivating conditional verifiable rewards as essential industrial components.}

\begin{document}

\maketitle

\section{Introduction}
\label{section:intro}

Recommendation systems have become indispensable across modern online platforms, mitigating information overload by surfacing items aligned with users' interests\citep{ramanujam2025large,behdin2025scaling,deng2025onerec}. Over the past decade, deep neural networks built on large embedding tables have dominated the modeling of user--item interactions~\citep{liang2025external,luo2025meta,zhang2024wukong,zhang2022dhen,wang2021dcn, zhou2019deep}.
More recently, Large Language Models (LLMs) have emerged as a transformative paradigm for recommendation~\citep{zhou2025efficiency,zhang2025reasonrec, zhao2024recommender,wu2024survey}, driven by their continual scaling gains, broad world knowledge, and contextual understanding. Representative efforts include P5~\citep{geng2022recommendation}, which unifies diverse recommendation tasks within a single LLM model; OneRec-Think~\citep{liu2025onerec}, which fuses retrieval and ranking stages via fine-tuned LLMs; and PLUM~\citep{he2025plum} that adapts pre-trained LLMs for scalable retrieval at YouTube.

Within the multi-stage industrial funnel, the final \emph{re-ranking} step disproportionately shapes what users see and interact with---most notably in carousel and grid layouts where top positions dominate engagement~\citep{gao2025llm4rerank,gao2024smlp4rec,liuneural}. Despite this, LLM-based re-ranking remains underexplored: existing work concentrates on retrieval and early ranking, while industrial re-rankers still rely on point-wise CTR scoring without explicit reasoning over user intent or item semantics. Three concrete gaps block adoption at scale.
\textbf{(G1) Reasoning is left on the table.} LLMs are typically deployed zero-shot or supervised-fine-tuned on plain ranking labels, leaving the chain-of-thought capabilities most effectively elicited through reinforcement learning (RL) on verifiable rewards largely unused for re-ranking.
\textbf{(G2) Vocabulary mismatch.} Industrial catalogs index billions of items via non-semantic identifiers that lie outside any base-LLM vocabulary, breaking the model's ability to reason about candidates directly.
\textbf{(G3) Industrial-scale tax.} Naive recipes incur prohibitive training cost, low serving throughput from long thinking traces, and---as we show---reward hacking that inflates offline metrics without reflecting genuine ranking quality.

\begin{figure*}[htbp]
  \centering
  \includegraphics[width=\linewidth]{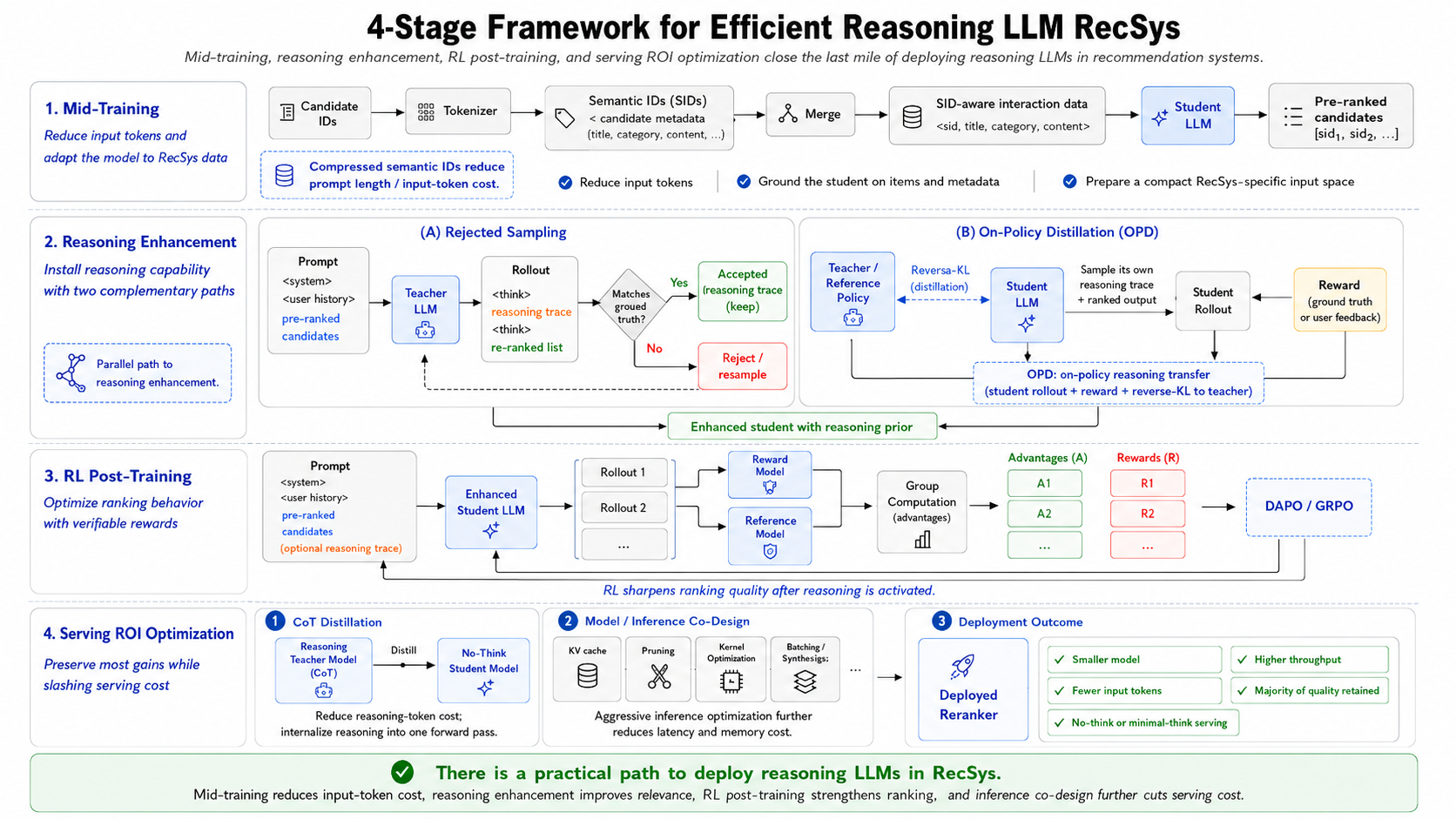}
  \caption{Overview of the 4-stage training pipeline: student LLM mid-training on tokenized semantic IDs, reasoning data generation using teacher LLM via rejection sampling and on-policy distillation, student LLM reasoning enablement by RL, and serving ROI optimization.}
  \label{fig:framework}
\end{figure*}

To close these gaps we introduce \approach, the \emph{Generative Reasoning Re-Ranker}. \approach is built on a three-stage training pipeline (Figure~\ref{fig:framework}) tailored to large-scale re-ranking. \emph{First}, a pre-trained student LLM (e.g., Qwen3-8B~\citep{qwen3technicalreport}) is mid-trained on semantic item IDs produced by a tokenizer that achieves $\geq$99\% uniqueness, allowing the model to recognize and generalize over the catalog without vocabulary blowup. \emph{Second}, a stronger teacher (e.g., Qwen3-32B~\citep{qwen3technicalreport}) is prompted with re-ranking-specific templates to generate hierarchical reasoning traces; targeted sampling or rejection sampling yields a curated SFT corpus that grounds the student's reasoning in user behavior and item semantics. \emph{Third}, we adapt DAPO~\citep{yu2025dapo} with a re-ranking-specific reward composed of a format term, an AUC/NDCG-based verifiable ranking term, and an optional LLM-as-a-judge reasoning term, providing scalable, reward-driven supervision.
Bridging this recipe to industry exposes failure modes that is often overlooked. Our contribution is to propose targeted solutions for each: a \emph{context compressor} that cuts input length by ${>}80\%$ while matching full-context quality, amortizing training cost; \emph{On-Policy Distillation (OPD)} as a scalable replacement for SFT---naive SFT collapses at industrial scale, whereas a 1.7B OPD student recovers $\approx$82\% of a 32B teacher's gain at 5\% of its size; \emph{de-hacking conditional reward}, without which RL learns to preserve the incoming order or fail to generate valid output format; and \emph{reasoning distillation} that internalizes chain-of-thought into a non-thinking policy, yielding ${\sim}15\times$ serving ROI at iso-quality.

We summarize our contributions as follows:
\begin{itemize}[leftmargin=*,nosep]
\item \textbf{Re-ranking-first LLM design.} Distinct from prior work that adapts LLMs for retrieval and early ranking, we establish design principles for LLM-based re-ranking and demonstrate that LLM world knowledge and reasoning translate into measurable ranking gains.
\item \textbf{Semantic-ID mid-training.} A tokenization scheme that converts non-semantic item IDs into semantic IDs with $\geq$99\% uniqueness, combined with a mixed mid-training corpus of semantic IDs and world knowledge, enabling LLMs to reason over catalog items directly.
\item \textbf{Reasoning activation pipeline.} Re-ranking-specific prompts paired with targeted and rejection sampling produce hierarchical reasoning traces; SFT (or OPD at industrial scale) on these curated traces installs a reasoning prior that pure RL cannot recover.
\item \textbf{Verifiable-reward RL for re-ranking.} A re-ranking-tailored adaptation of DAPO with a multi-component reward; we identify and mitigate two reward-hacking modes (incoming-order preservation and position-bias exploitation) through conditional verifiable rewards and debiased training data.
\item \textbf{Industrial adaptations.} A context compressor, OPD-based scalable distillation, and CoT distillation for non-thinking serving---each ablated to quantify its contribution to deployment feasibility.
\item \textbf{Comprehensive evaluation.} \approach improves over internal baselines on industrial traffic (+18.7\% R@1, +9.6\% N@3) and shows no decay over subsequent 9 days serving; ablations confirm that OPD supplies the reasoning prior and RL sharpens the ranking objective on top of it.
\end{itemize}

\section{Tokenized Mid-Training}\label{sec:stage_1}
In this section, we introduce the key components for attaining the LLM-based generative recommender: \textbf{tokenization} and \textbf{mid-training}. 
To obtain the generative retriever, we stemmed from the mid-training workflow from the advanced LLM-based generative retriever that perform multi-task training on an LLM.

\subsection{Tokenizer and Semantic ID (SID)}

Learning the so-called semantic IDs (SIDs) has been widely adopted in recommendation systems to mitigate generalization issues pertaining to large embedding tables. 
The seminal work, TIGER~\citep{rajput2023recommender}, was the first to apply this technique to sequential recommendation. 
At a high level, given the textual feature $x$ of an item, the tokenizer maps it to a sequence of discrete integers, i.e., a compact symbolic representation, defined as $\text{Tokenizer}(x) = (z_1, z_2, \ldots, z_K)$ where the final representation is a memeber of the set $\{1,\ldots,C_1\}\times \cdot\cdot\cdot \times \{1,\ldots,C_K\}$,
where $C_i$ is the cardinality of the $i$-th codebook. The core of the tokenizer is an (RQ-VAE)~\citep{lee2022autoregressive}. The SID tokens are added to the vocabulary as special tokens.
We refer the reader to Sec 2 from our previous technical paper~\citep{liang2026generative} for further details.

\subsection{Mid-Training via Multi-Task Learning}
\label{sec:item alignment}

TIGER~\citep{rajput2023recommender} trains an autoregressive model purely over Semantic ID sequences. In the era of LLMs, leveraging world knowledge encoded in pretrained LMs has become an important direction. 
To this end, OneRec-Think~\citep{liu2025onerec} developed by Kuaishou introduces a mid-training stage. Referred to as the `item alignment', this stage enables the LLM to align the recommendation knowledge and language (Semantic IDs) with the LLM's linguistic space and world knowledge. More concretely, the idea here is to \emph{interleave Semantic IDs (SIDs) with natural language tokens within a single sequence} and to optimize the SID embedding table through the \emph{next-token prediction objective}. 
The mid-training tasks follows the same design from our previous technical paper~\citep{liang2026generative}.

\section{Reasoning Enhancement}\label{sec:stage_2}
We now present the design of reasoning enhancement for re-ranking. In Sec.\ref{sec:chat_template}, we introduce a chat-format training-sample structure that grounds item representation in semantic IDs and enables chain-of-thought reasoning with structured JSON output. In Sec.\ref{Reasoning Trace Generation}, we present two complementary strategies for generating high-quality reasoning traces, i.e., \emph{targeted sampling}, which leverages ground-truth guidance, and \emph{rejection sampling}, which ensures reasoning authenticity through iterative verification, together with five prompt-design principles that enable grounded, interpretable, and behaviorally-aware reasoning from large language models.

\subsection{Chat-formatted Template of Reasoning Data}\label{sec:chat_template}
We construct training samples in a chat-format structure with three message roles. The design follows six principles:
\begin{itemize}[leftmargin=*,nosep]
  \item \textbf{Role-based System Prompt.} The system message instantiates an analyst persona and specifies the re-ranking objective, activating domain reasoning without invoking purchase-intent or persuasion framing.
  \item \textbf{Rich Item Metadata.} Each item carries its title and category hierarchy, supporting both semantic and category-aware reasoning.
  \item \textbf{Unified Item Format.} History and candidates share the same structural format (SID + title + categories), ensuring consistent representation learning across input contexts.
  \item \textbf{Chain-of-Thought Reasoning Trace.} The assistant response contains step-by-step reasoning that explicitly cites items by their SIDs, producing grounded, verifiable traces.
  \item \textbf{Structured JSON Output.} The response is a JSON object containing the reasoning explanation and the ranked list, enabling deterministic parsing and evaluation.
\end{itemize}

\subsection{Reasoning Trace Generation}
\label{Reasoning Trace Generation}

To generate high-quality reasoning trace, we consider two generation strategies for the reasoning trace: \emph{targeted sampling} and \emph{rejection sampling}.

\paragraph{Targeted Sampling.}
In the targeted approach, we provide the target item $\mathbf{s}_{v_{n+1}}$ and most recent $k$ historical items, then query a larger LLM to generate a reasoning trace $\tau$ explaining the target interaction:
\begin{equation}
    \tau \sim P_\theta(\cdot | \mathcal{P}_{\text{targeted}} ([\mathbf{s}_{v_1},...,\mathbf{s}_{v_k}], [\mathbf{s}_{y_1},...,\mathbf{s}_{y_c}], \mathbf{s}_{v_{n+1}})),
\end{equation}
where $\mathcal{P}_{\text{targeted}}(x, y, z)$ constructs a prompt to query the rationale for why a user who interacts with item sequence $x$ would be most interested in $z$ from candidate list $y$. 
As the ground truth is involved, the targeted approach always yields rationales for why the target might be favored by the user. Notice that this requires having ground truth in the prompt. 

\paragraph{Rejection Sampling.}
In sharp contrast to targeted sampling, in rejection sampling we do not provide the ground truth. Instead we repeatedly query the LLM to predict which item $\mathbf{\hat{s}}_{y_c}$ among the pre-ranked candidates is most likely to be the user's next interest, until the prediction matches the ground-truth target:
\begin{equation}
\begin{split}
(\tau, \hat{\mathbf{s}}_{y_c}) \;\sim\;& P_\theta\!\left(\cdot \,\middle|\, \mathcal{P}_{\text{rejection}}\!\bigl([\mathbf{s}_{v_1},\ldots,\mathbf{s}_{v_k}],\,[\mathbf{s}_{y_1},\ldots,\mathbf{s}_{y_c}]\bigr)\right) \\
& \text{s.t.}\quad \hat{\mathbf{s}}_{y_c} = \mathbf{s}_{v_{n+1}}.
\end{split}
\end{equation}
where $\mathcal{P}_{\text{rejection}}(x,y)$ constructs a prompt to query which item in candidate list $y$ is most likely to be the user's next interest, given history $x$.

To generate high quality reasoning trace for re-ranking, we leverage following principles to design the prompt for instructing LLM for reasoning:
\begin{itemize}[leftmargin=*,nosep]
    \item \textbf{Concrete System Role and Re-ranking Task definition}: To incentivize re-ranking capabilities, we elucidate the recommendation domain, the dedicated roles for the LLM to display, and the concrete definition of the re-ranking task.
    \item \textbf{Collaborative Context Presentation}: To provide comprehensive decision context, we present both the user's engagement history (with structured metadata including titles and categories) and the complete candidate set, enabling the model to perform holistic comparison rather than isolated item evaluation.
    \item \textbf{Domain Knowledge Priming}: To leverage sequential engagement patterns inherent in e-commerce (e.g., shampoo $\rightarrow$ conditioner $\rightarrow$ styling products), we explicitly prompt the model to consider such domain-specific heuristics, enabling it to apply common-sense reasoning about product complementarity and routine-based purchasing behavior.
    \item \textbf{Critical Guidelines as Output Constraint}: To ensure reasoning traces are grounded and verifiable, we impose explicit constraints such as requiring the model to cite items by their SIDs. This forces the model to anchor its reasoning in specific historical items rather than generating hallucinated justifications, enabling direct traceability between reasoning steps and context.
    \item \textbf{Structured Multi-Step Reasoning Format}: To elicit progressive and interpretable reasoning, we provide an explicit step-by-step output format with examples. This hierarchical structure guides the model through: (1) broad pattern recognition, (2) identifying complementary product types, and (3) matching to specific candidates, mirroring human decision-making processes.
\end{itemize}

\subsection{Supervised Fine-tuning on Reasoning Traces}
\label{sec:sft_formulation}

\paragraph{Problem Setup and Prompt Interface}
Given a user engagement history and a pre-ranked list of candidate items produced by a previous stage, our desire is to re-rank the candidates so as to promote the ground-truth next item.
Each training or inference instance is formatted as a chat prompt consisting of three roles:
\texttt{system}, \texttt{user}, and \texttt{assistant}. 
During supervised fine-tuning (SFT), the model is trained to generate the full assistant message.
During reinforcement learning (RL), the policy conditions only on the system and user messages, and the assistant output is treated as the action.

Recommender models often struggle to generate effective CoT reasoning from noisy and lengthy real-world user behavior sequences~\citep{liu2025onerec}. The generated reasoning traces are used to supervise a base LLM to acquire reasoning capability for re-ranking. We fine-tune the model to generate both the reasoning trace $\tau = [r_1, \ldots, r_M]$ and the ranked output $\mathbf{o} = [o_1, \ldots, o_T]$.

To preserve ranking performance while enabling reasoning, we decouple the losses for reasoning and ranking tokens. Specifically, we apply the language modeling loss \emph{only to the assistant message}, with separate weights for reasoning and ranking segments:
\begin{equation}
\mathcal{L}_{\text{SFT}} =
- \lambda_r \sum_{i=1}^{M}
\log P(r_i \mid \mathcal{P}, r_{<i})
- \lambda_o \sum_{j=1}^{T}
\log P(o_j \mid \mathcal{P}, \tau, o_{<j}),
\end{equation}
where $\lambda_r < \lambda_o$ balances reasoning fluency and ranking accuracy. This training procedure teaches the model to generate grounded, coherent reasoning traces that connect user history with candidate comparisons, while maintaining strong re-ranking behavior.

\subsection{OPD for Reasoning Activation}
\label{sec:method_distill}
We use \emph{on-policy distillation} (OPD)~\citep{zhao2026self} to transfer reasoning from a frozen teacher into the deployed student. A GRPO-style loop treats the student as the trainable actor and the teacher as the reference policy: at every step the student samples reasoning chains and ranked lists from its own distribution $\pi_\theta$, each rollout receives the reward of Section~\ref{sec:stage_3}, and the actor is updated with a clipped surrogate plus a per-token reverse-KL anchor to the teacher,
\begin{equation}
\begin{aligned}
\mathcal{L}_{\mathrm{OPD}}(\theta)
={}& -\mathbb{E}\!\left[\min\!\bigl(\rho_t \hat A_t,\;\mathrm{clip}(\rho_t,1{-}\epsilon_{\mathrm{lo}},1{+}\epsilon_{\mathrm{hi}})\hat A_t\bigr)\right] \\
& {}+ \beta\,\mathbb{KL}\!\bigl[\pi_{\theta}(\cdot\!\mid\! s_t)\,\|\,\pi_{\mathrm{T}}(\cdot\!\mid\! s_t)\bigr],
\end{aligned}
\label{eq:opd}
\end{equation}
where $\rho_t$ is the per-token importance ratio, $\hat A_t$ the group-relative advantage, and $\beta$ the distillation strength. The teacher contributes only token log-probabilities, never labels, acting as a distributional anchor while gradients flow through the student.

\paragraph{Why on-policy beats off-policy CoT supervision.}
The two CoT-SFT baselines introduced earlier in Section~\ref{sec:sft_formulation} both reduce
reasoning transfer to behavior cloning on a static, teacher-defined corpus, and inherit the standard pathologies of teacher-forced sequence learning: the student is supervised on states it will never visit at deployment, producing a train/inference distribution mismatch that compounds along the generated chain. Each method also adds its own bias on top of this. Targeted sampling conditions the teacher on the answer, so the resulting traces tend to encode post-hoc, label-aware shortcuts that the student memorizes as fluent but non-causal reasoning. Rejection sampling, by contrast, silently discards every prompt on which the teacher cannot meet the AUC bar within the sampling budget, which both wastes teacher compute and removes precisely the hardest examples—where supervision is most needed—from the training set. OPD avoids both failure modes by construction: the gradient is computed on trajectories the student itself samples, every prompt contributes a continuous reward signal regardless of how easy or hard it is for the teacher, and the teacher's role is reduced from generating the target trace to regularizing the student's own distribution. This combination
of on-policy exposure, dense per-rollout reward, and per-token KL anchoring is what allows the small student to absorb the teacher's reasoning competence without inheriting its trace-style mismatch.

\section{RL Post-Training}\label{sec:stage_3}

While SFT enables coherent reasoning, it does not directly optimize the re-ranking objective. Building on the distilled reasoning capability of Sec.~\ref{sec:sft_formulation}, we apply RL to refine both the reasoning process and the final ranking quality. Given a prompt $\mathcal{P}(\mathcal{H},\mathcal{D})$, the policy $\pi_\theta$ generates an output $o=(\tau,\mathbf{o})$; during RL the policy conditions only on the system and user messages and $o$ is treated as the action. Unlike public benchmarks, where each impression has a single positive next item, industrial slates are \emph{multi-positive}: a candidate list may carry several engagement labels at once. We therefore replace the single-target rank-delta of prior work with reward functions that score the whole permutation against a multi-label ground truth.

\subsection{Ranking Reward}\label{sec:rank_reward}
A candidate slate $\mathcal{D}=\{c_1,\ldots,c_K\}$ is associated with a binary engagement label vector $\mathbf{y}\in\{0,1\}^K$ that may contain multiple ones. We adopt the per-impression AUC of the predicted permutation $\pi$ against $\mathbf{y}$ as the primary ranking signal:
\begin{equation}
R_{\text{AUC}}(\pi,\mathbf{y}) \;=\; \frac{1}{|\mathcal{M}||\mathcal{N}|}\sum_{i\in\mathcal{M}}\sum_{j\in\mathcal{N}}\mathbf{1}\!\left[\mathrm{rank}_\pi(i)<\mathrm{rank}_\pi(j)\right],
\end{equation}
where $\mathcal{M}=\{i|y_i{=}1\}$ and $\mathcal{N}=\{j|y_j{=}0\}$. AUC handles multiple positives natively, is bounded in $[0,1]$, and is permutation-invariant within each label class, providing a stable RL signal.

When richer post-engagement signals are available, we further incorporate a graded-relevance \emph{NDCG} term using a three-level label $g_i\in\{0,1,2\}$ for $\{\text{no engagement},\text{ click},\text{ click}+\text{conversion}\}$:
\begin{equation}
R_{\text{NDCG}}(\pi,\mathbf{g}) \;=\; \frac{1}{Z}\sum_{i=1}^{K}\frac{2^{g_{\pi^{-1}(i)}}-1}{\log_2(i+1)},
\end{equation}
with $Z$ the ideal DCG. The combined ranking reward is $R_{\text{rank}}=R_{\text{AUC}}~\text{or}~R_{\text{NDCG}}$ in our internal-data setup. Slates with no positives or no negatives ($|\mathcal{P}|=0$ or $|\mathcal{N}|=0$) are filtered at data-loading.

\subsection{Conditional Format Reward}\label{sec:fmt_reward}
To ensure parseable outputs, we add a format reward $R_{\text{fmt}}=\Omega(o)$ that checks whether (i) the reasoning trace $\tau$ and ranking output $\mathbf{o}$ can be reliably parsed, and (ii) the parsed ranking is a valid permutation of $\{1,\ldots,K\}$ matching the slate size. Naively combining $R_{\text{rank}}$ and $R_{\text{fmt}}$ exposes the policy to two reward-hacking pathways. First, an invalid permutation could still earn a non-trivial $R_{\text{rank}}$ from a partial parse; we therefore gate $R_{\text{rank}}$ on $\Omega(o)=1$. Second, when the upstream slate is itself non-trivially ordered---i.e., the identity permutation already attains $R_{\text{rank}}<1$---the policy can harvest format reward by emitting $[1,2,\ldots,K]$ without performing any re-ranking, riding on whatever AUC the input order already supplies. We detect this \emph{identity-permutation cheating} and zero out $R_{\text{rank}}$ when the model output equals the input order \emph{and} the input order is suboptimal:
\begin{equation}
R =
\begin{cases}
R_{\text{rank}} + \alpha R_{\text{fmt}}, & \pi\neq[1,\ldots,K]\ \text{or}\ R_{\text{AUC}}([1,\ldots,K],\mathbf{y})=1,\\
\alpha R_{\text{fmt}}, & \pi=[1,\ldots,K]\ \text{and}\ R_{\text{AUC}}([1,\ldots,K],\mathbf{y})<1,
\end{cases}
\end{equation}
where $\alpha$ is the format-reward weight. When the upstream slate is already optimal, the identity permutation is the genuinely correct answer and reward flows normally.

\subsection{Training via DAPO}

We optimize the policy using Decoupled Clip and Dynamic sAmpling Policy Optimization (DAPO) algorithm~\citep{yu2025dapo}, a recently state-of-the-art RL algorithm developed upon Grouped Policy Optimization
(GRPO)~\citep{shao2024deepseekmath}. 
It can effectively resolve the entropy collapse phenomenon and rollout length bias identified in the GRPO training process. For each prompt, we sample a group of $G$ outputs
$\{o_i\}_{i=1}^G$
and optimize:
\begin{equation}
\begin{aligned}
\mathcal{J}_{\text{DAPO}}(\theta) = \ &\mathbb{E}_{(q,a)\sim\mathcal{D},\,\{o_i\}_{i=1}^G \sim \pi_{\theta_{\text{old}}}(\cdot|q)} \\
&\Bigg[ \frac{1}{\sum_{i=1}^G |o_i|} \sum_{i=1}^G \sum_{t=1}^{|o_i|} \min\Big( r_{i,t}(\theta)\,\hat{A}_{i,t}, \\
&\qquad \text{clip}\big(r_{i,t}(\theta),\, 1-\varepsilon_{\text{low}},\, 1+\varepsilon_{\text{high}}\big)\,\hat{A}_{i,t} \Big) \Bigg],
\end{aligned}
\label{eq:dapo}
\end{equation}
where
\begin{equation}
r_{i,t}(\theta) = \frac{\pi_\theta(o_{i,t} | q, o_{i,<t})}{\pi_{\theta_{\text{old}}}(o_{i,t} | q, o_{i,<t})}, \quad \hat{A}_{i,t} = \frac{R_i - \text{mean}(\{R_i\}_{i=1}^G)}{\text{std}(\{R_i\}_{i=1}^G)}.
\label{eq:9}
\end{equation}
$\pi_\theta$ and $\pi_{\theta_{\text{old}}}$ denote the current and previous policies, respectively. 
The pair $(q,D)\sim\mathcal{D}$ is sampled from the training distribution. The advantage estimate $\hat{A}_{i,t}$ is computed for each output sequence $o_i$ and normalized within the group $G$. 
Adhering to the Clip-Higher strategy, DAPO decouples the lower and higher clipping range as $\varepsilon_{\text{low}}$ and $\varepsilon_{\text{high}})$. 
Additionally, to avoid zero policy gradients (advantages) and improve sample efficiency, it oversamples and filters out prompts with the accuracy equal to 1 and 0.

\section{Serving ROI Optimization}

The first three stages yield a capable but expensive reasoning ranker:
mid-training adapts the model to RecSys inputs and compresses prompts via
semantic IDs (\S\ref{sec:stage_1}), reasoning enablement instills a reasoning
prior (\S\ref{sec:stage_2}), and RL post-training sharpens ranking behavior
under that prior (\S\ref{sec:stage_3}). The final stage preserves these ROI
gains while slashing serving cost along its two dominant axes: the
\emph{input cost} of prompting over an $O(B)$ item catalog, and the
\emph{decode cost} of generating a chain-of-thought (CoT) per request. We
address them with two complementary levers---input-side context compression
and decode-side reasoning internalization---delivering higher throughput and
lower latency without sacrificing ranking quality.

\subsection{Input-Side: Context Compression}

Mid-training over an $O(B)$ item catalog requires generating large volumes of
SID-aligned data and incurs heavy GPU cost per iteration, which slows online
experimentation. As a faster alternative for cutting input cost, we train a
\emph{context compressor} that shrinks the input before it reaches the ranker.

The compressor is trained with GRPO using an LLM-as-a-judge that scores each
compressed output along three axes: \textit{solvability} $s\!\in\!\{0,1\}$,
\textit{information preservation} $p\!\in\!\{1,\ldots,10\}$, and
\textit{ranking quality} $q\!\in\!\{1,\ldots,10\}$. The judge prompt is
mode-aware (item-level vs.\ free-form summary). The reward combines a
continuous compression term with a solvability-conditioned judge term:
\begin{equation}
    r_{\text{judge}} =
    \begin{cases}
        0.2\,\bar{p} + 0.8\,\bar{q} & s = 1 \\
        0.8\,\bar{p} + 0.2\,\bar{q} & \text{otherwise}
    \end{cases}
\end{equation}
\begin{equation}
    r = \bigl(\alpha\, r_{\text{comp}} + (1{-}\alpha)\, r_{\text{judge}}\bigr) \cdot \lambda_{\text{ellipsis}},
\end{equation}
where $\bar{p}=p/10$, $\bar{q}=q/10$,
$r_{\text{comp}}=\max(0,\,1-|\text{compressed}|/|\text{original}|)$, $\alpha$
trades off compression against quality, and
$\lambda_{\text{ellipsis}}\!\in\![0,1]$ penalizes truncation shortcuts. When
the task is solvable we weight ranking signal over raw information
completeness, since we find preservation correlates weakly with downstream
ranking while higher compression can suppress noise.

\subsection{Decode-Side: Internalizing Reasoning for Efficient Serving}

Generating a full CoT per request is the dominant \emph{decode} cost of GR2.
To preserve the reasoning gains under large-scale serving latency, we run a
second RL pass on the RL post-training checkpoint (\S\ref{sec:stage_3}) with
explicit reasoning bypassed: the policy is trained to emit only the structured
ranked output, conditioned directly on the system and user prompts. Because
the reasoning prior---instilled via OPD during reasoning enablement
(\S\ref{sec:stage_2}) and sharpened by RL (\S\ref{sec:stage_3})---is already
encoded in the shared backbone, it transfers to the direct-output policy
without incurring CoT generation cost. This reasoning-free variant is the
deployed artifact; the CoT-enabled model is retained offline for evaluation
and reward design.

\subsection{Systems-Level Optimizations: Pruning and KV Caching}

The above levers reduce token counts; two systems-level techniques further cut
the cost per token. \emph{Model pruning} shrinks the served network via
depth-wise (layer) pruning followed by redistillation to recover quality,
lowering compute across both prefill and decode. \emph{KV caching} exploits the
structure of RecSys prompts: the candidate set for a request is fixed and
largely shared across users, so we precompute its key--value representations
once and reuse them across requests, ordering the prompt as system prompt,
candidates, then user context so the candidate KV remains reusable. Together
these amortize compute over the stable parts of the input and the reduced
model, compounding the savings from compression and reasoning internalization.

\section{Experiments}\label{sec:exp}
\label{experiments}

We focus this section on the industrial setting and answer the following research questions:

\begin{itemize}[leftmargin=*,nosep]
    \item \textbf{Q1}: Does GR2 outperform the point-wise lagacy baseline, and is the gain robust to test-set scale, traffic freshness, and student model size?
    \item \textbf{Q2}: How do different reasoning-activation strategies (SFT, OPD) and RL post-training affect re-ranking quality at industrial scale?
    \item \textbf{Q3}: Can the enterprise-scale adaptations---context compression, OPD-based distillation, and implicit-CoT serving---preserve ranking quality under industrial latency and GPU budgets?
\end{itemize}

Public-benchmark results, the semantic-ID tokenizer design, and the codebook/uniqueness study are reported in a prior technical paper~\cite{liang2026generative}; here we concentrate on the industrial setting, where adaptation challenges dominate.

\subsection{Experiment Setup}

\subsubsection{\textbf{Datasets}}
\label{sec:datasets}

GR2 is trained on a single day (\texttt{01-25}) of internal logs containing roughly 70k user sessions and evaluated on held-out sessions from \texttt{02-01} through \texttt{02-09}.
Each training example corresponds to one impression list paired with the user's historical feedback (e.g., clicked products). We apply standard industrial filtering to drop duplicated logging events and users with fewer than three interactions in the observation window.
To stress-test generalization, the test split is constructed so that user identifiers and interacted products rarely overlap with training: $>$99\% of candidate products, 100\% of user IDs, and 93\% of user-history items are unseen during training. User histories are aggregated up to day~$-2$ relative to the impression to prevent label leakage.
Unless noted otherwise, we report relative gains over the point-wise legacy baseline, which is online-trained with snapshot refresh every 60--90 minutes.

\subsubsection{\textbf{Metrics}}
\label{sec:metrics}

\begin{figure}
    \centering
    \includegraphics[width=0.75\linewidth]{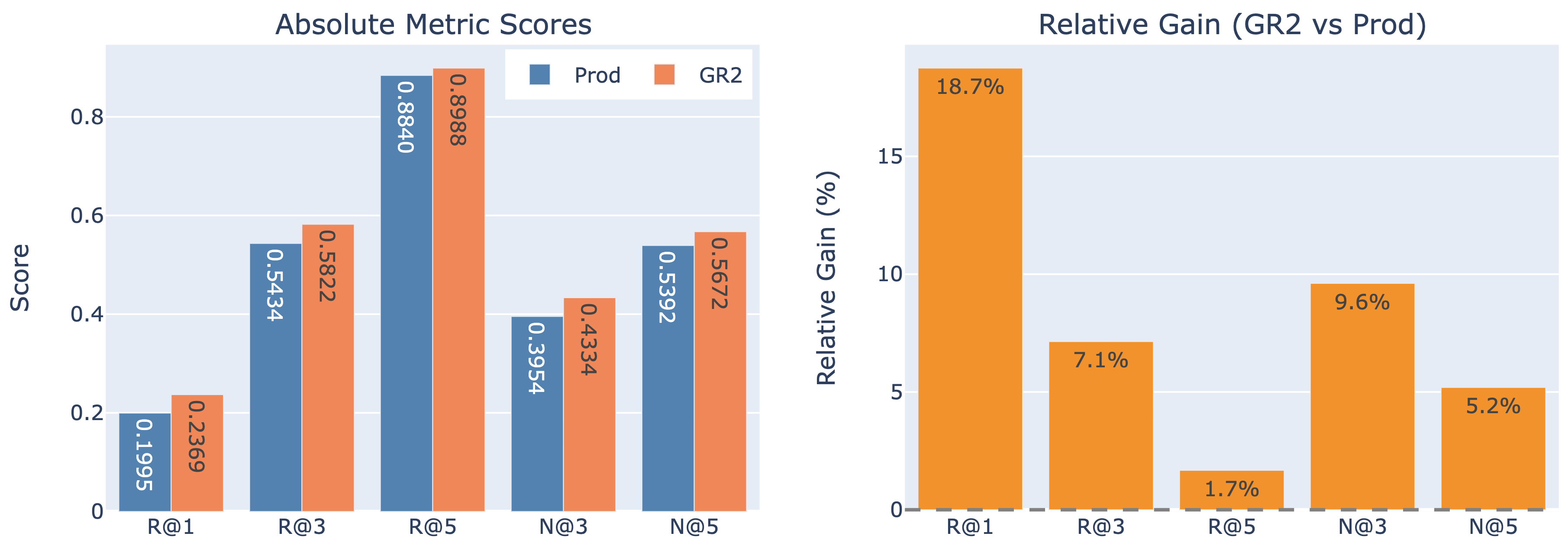}
    \caption{Re-ranking performance comparison with legacy baseline on industrial traffic across 70k sessions.}
    \label{fig:main_results}
\end{figure}

We report standard sequential-recommendation metrics, \emph{Recall@K} and \emph{Normalized Discounted Cumulative Gain at K (NDCG@K)}, computed over held-out impression lists.

\subsection{Main Results}
\label{sec:main_results}

\paragraph{GR2 substantially outperforms the legacy baseline (Fig.~\ref{fig:main_results}).}
On the internal test set, GR2 delivers \textbf{+18.7\% R@1} and \textbf{+7.1\% R@3} over a heavily-tuned legacy baseline, despite being trained on a single day of logs.
Since the GR2 checkpoint is roughly two weeks stale relative to the test window while the baseline is refreshed hourly, the lift cannot be attributed to data freshness; it reflects a genuine improvement in ranking quality driven by reasoning over product semantics.

\begin{figure}
    \centering
    \includegraphics[width=0.75\linewidth]{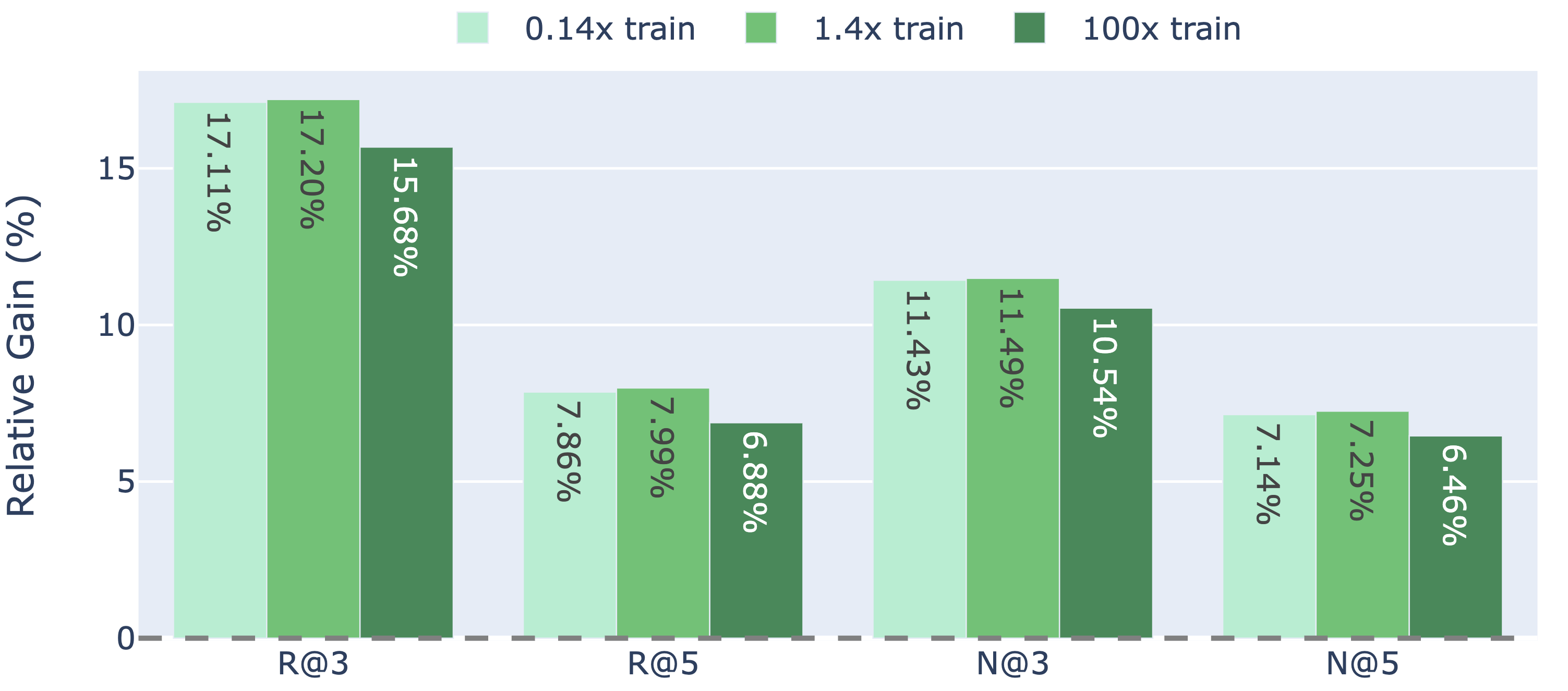}
    \caption{Relative performance gain across different testing data scales (regarding the training data size).}
    \label{fig:test_size_scaling}
\end{figure}

\paragraph{The lift is preserved at industrial scale (Fig.~\ref{fig:test_size_scaling}).}
Holding training data fixed at 70k sessions, we sweep test-set size from $0.14\times$ through $1.4\times$ up to $100\times$ training data size.
Relative R@K and N@K gains remain essentially constant across all three scales, indicating that GR2's improvement is not an artifact of small-sample evaluation and transfers to traffic volumes seen in industrial serving.

\begin{figure}
    \centering
    \includegraphics[width=0.75\linewidth]{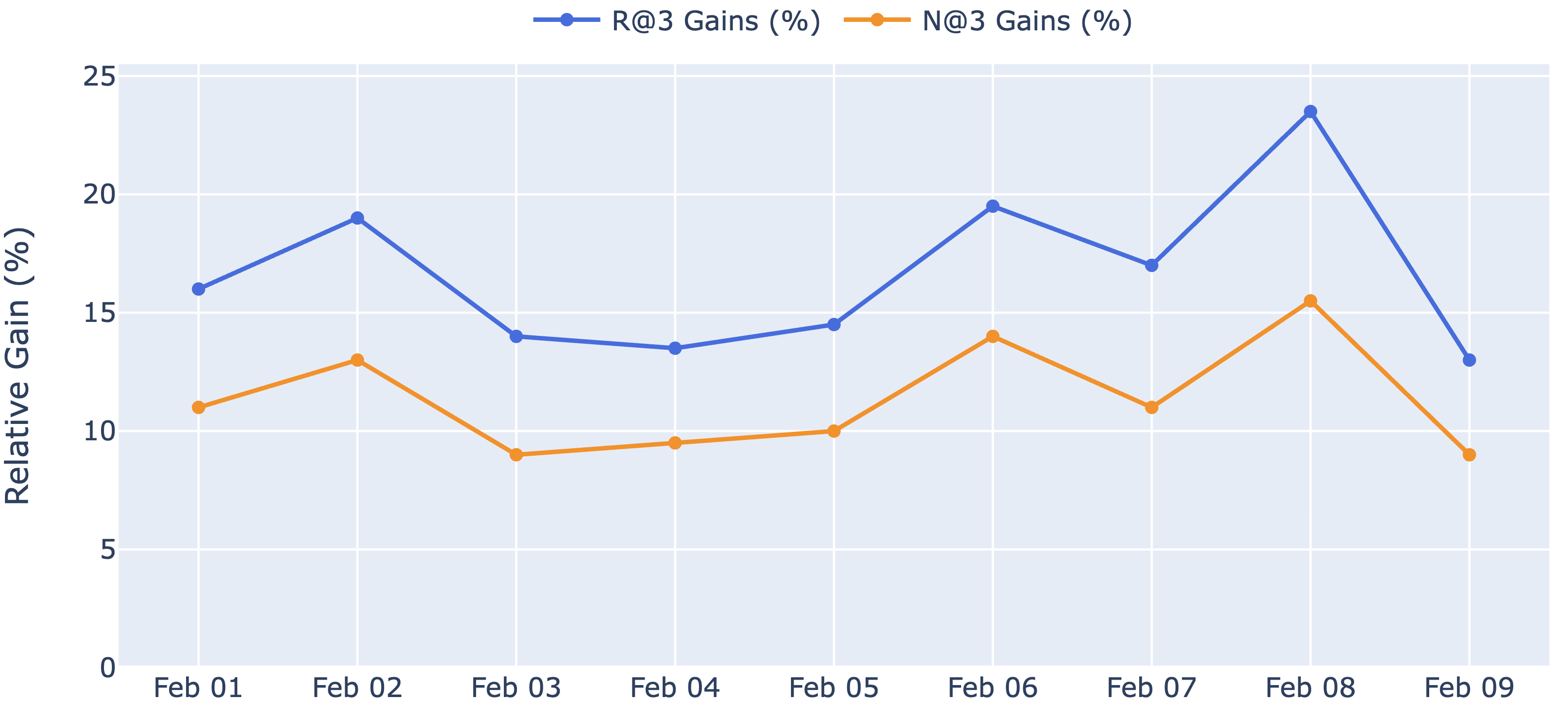}
    \caption{Relative performance gain tested on 9 consecutive days of traffic after 2 weeks training.}
    \label{fig:freshness}
\end{figure}

\paragraph{Gains do not decay under model staleness (Fig.~\ref{fig:freshness}).}
Evaluated daily for nine consecutive days starting \texttt{02-01}, the relative lift over the legacy baseline shows no measurable decay even though the GR2 checkpoint is more than two weeks stale by the end of the window.
This is striking given the cold-start setting described above (Section~\ref{sec:datasets}): nearly every user and product at test time is novel, yet GR2 generalizes without retraining.
We attribute this robustness to GR2's reliance on LLM world knowledge and reasoning rather than memorized sparse IDs, which is precisely what forces the legacy baseline to refresh every 60--90 minutes to avoid distribution drift.
We expect this gap to widen on lower-funnel signals (delayed conversions), where sparse-ID models are further disadvantaged by feedback delay and label sparsity.

\begin{figure}
    \centering
    \includegraphics[width=0.75\linewidth]{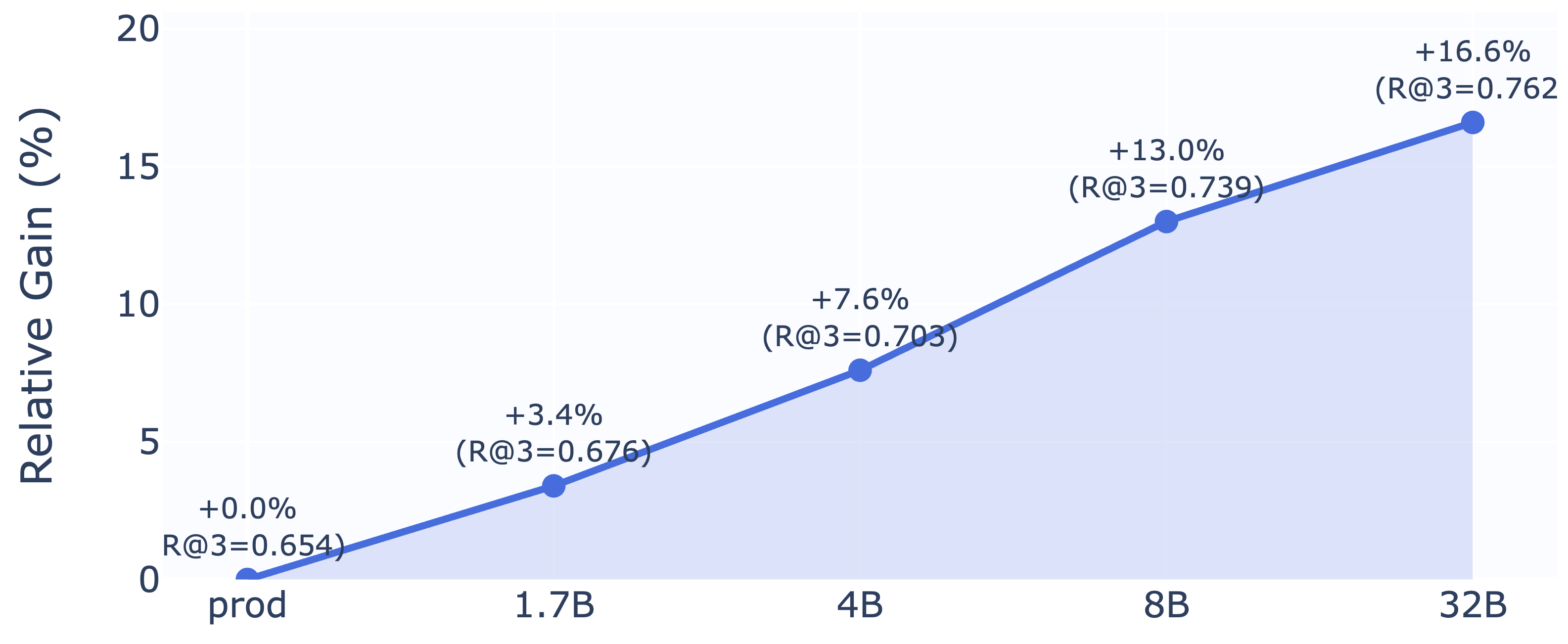}
    \caption{Relative R@3 gain for various Qwen3 model sizes.}
    \label{fig:model_scaling}
\end{figure}

\paragraph{Performance scales with model size (Fig.~\ref{fig:model_scaling}).}
Sweeping over Qwen-3 from 1.7B to 32B parameters, we observe R@3 lift grows monotonically with model size, demonstrating that GR2 inherits the LLM scaling law in the ads re-ranking domain.
This stands in contrast to sparse-ID baseline models, whose scaling is bounded by ID-vocabulary capacity rather than parameter count.
The current curve has not yet flattened despite our $\sim$10-GPU budget, suggesting potential headroom; we discuss how on-policy distillation (Section~\ref{sec:method_distill}) closes the resulting serving-cost gap.

\subsection{Serving Latency and Distillation}
\begin{figure}
    \centering
    \includegraphics[width=0.8\linewidth]{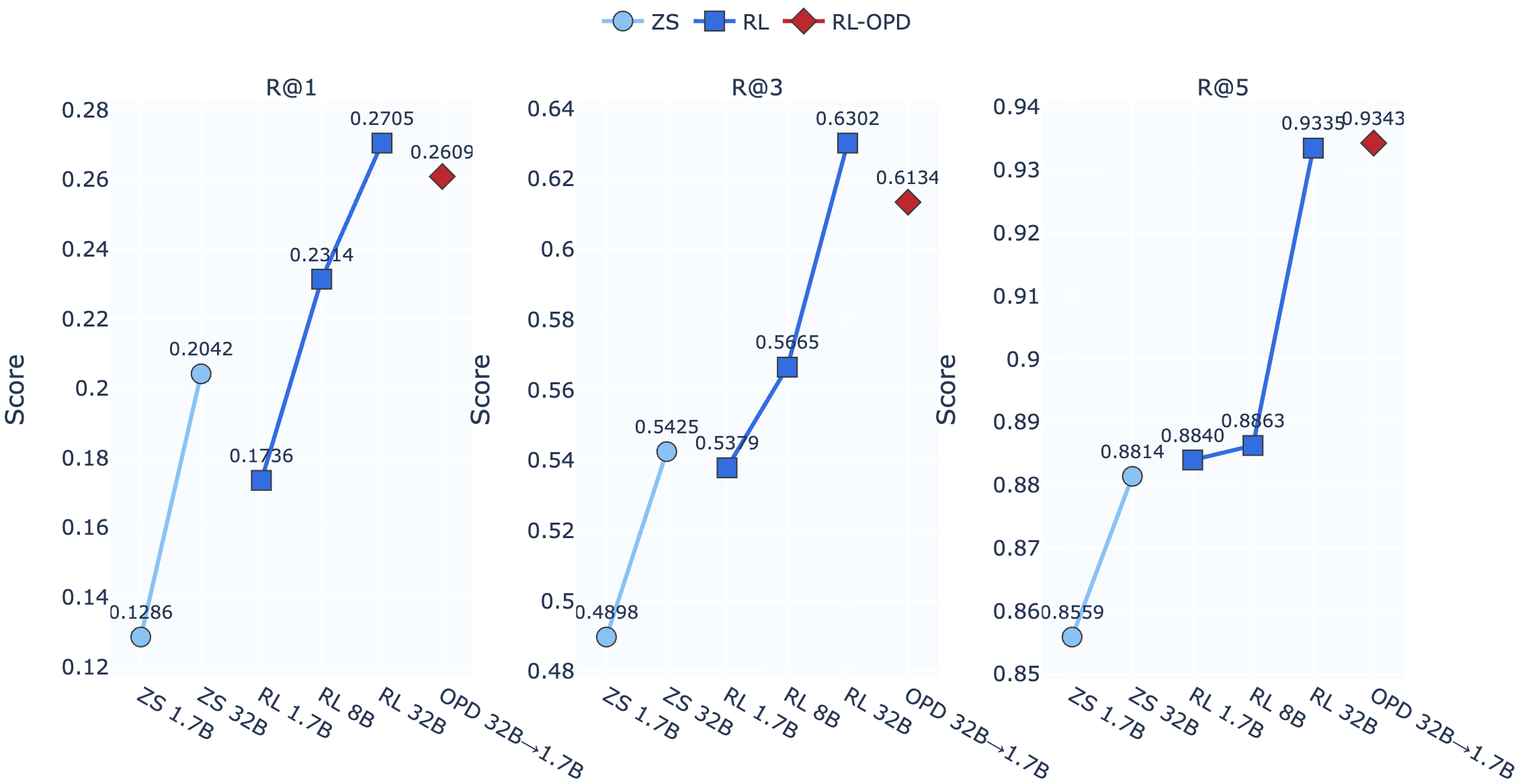}
    \caption{On-policy Distillation (OPD) can significantly improve Serving ROI. ZS denotes zeroshot.}
    \label{fig:opd_roi}
\end{figure}

\paragraph{Serving ROI via On-Policy Distillation (Fig.~\ref{fig:opd_roi})}
Our default GR2 recipe, denoted \textbf{RL-OPD}, first applies OPD to bootstrap reasoning without the catastrophic forgetting we observe with token-level SFT, then runs RL post-training on the OPD checkpoint to sharpen ranking quality while preserving the reasoning prior.
A 1.7B student distilled from a 32B teacher (5\% of teacher size) recovers \textbf{82\%} of the 32B's gain over the baseline baseline and delivers \textbf{2.6$\times$} the gain of an 8B model trained without distillation, yielding a $\sim$\textbf{15$\times$} serving-ROI boost ($32/1.7 \times 0.82$) at iso-quality.

\begin{figure}
    \centering
    \includegraphics[width=0.75\linewidth]{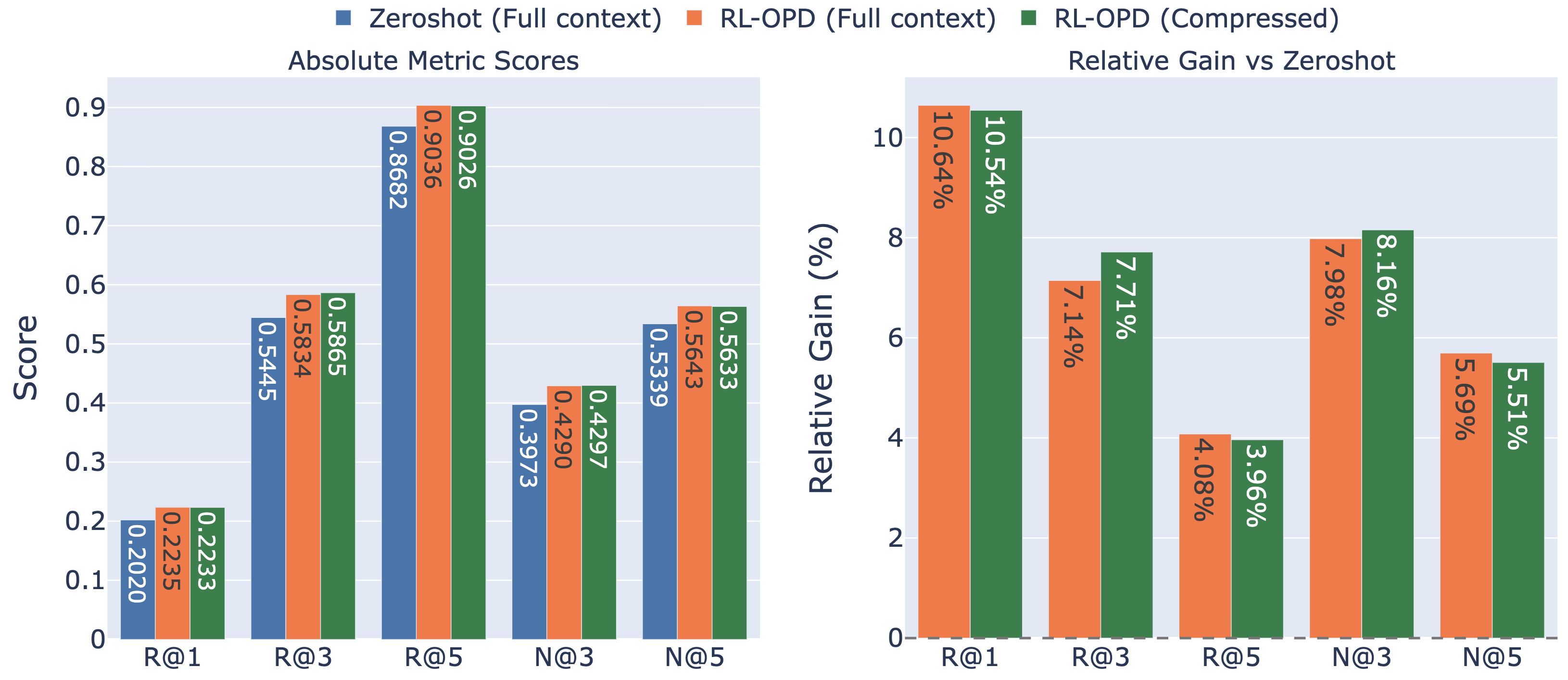}
    \caption{Performance Comparison between Full Context vs Compressed Context (80\% fewer tokens).}
    \label{fig:compressor}
\end{figure}

\paragraph{Context Compression (Fig.~\ref{fig:compressor})} 
Both RL-OPD variants improve substantially over the zero-shot 0.6B baseline (\textbf{+10--12\%} R@1, \textbf{+4--6\%} R@5/N@5). The compressed-context variant matches or slightly surpasses the full-context counterpart at R@3/N@3 (\textbf{+7.71\%} vs.\ \textbf{+7.14\%}; \textbf{+8.16\%} vs.\ \textbf{+7.98\%}) while consuming \textbf{$<$20\%} of the tokens---an additional \textbf{$>$5$\times$} context reduction at iso-quality. We attribute this to the solvability-conditioned reward: when $s{=}1$, the compressor is pushed toward ranking quality $\bar{q}$ rather than literal preservation $\bar{p}$, learning to discard ranking-irrelevant signal. This confirms our hypothesis that information preservation correlates only weakly with ranking utility, and that aggressive ranking-aware compression acts as a denoiser.

\begin{figure}
    \centering
    \includegraphics[width=\linewidth]{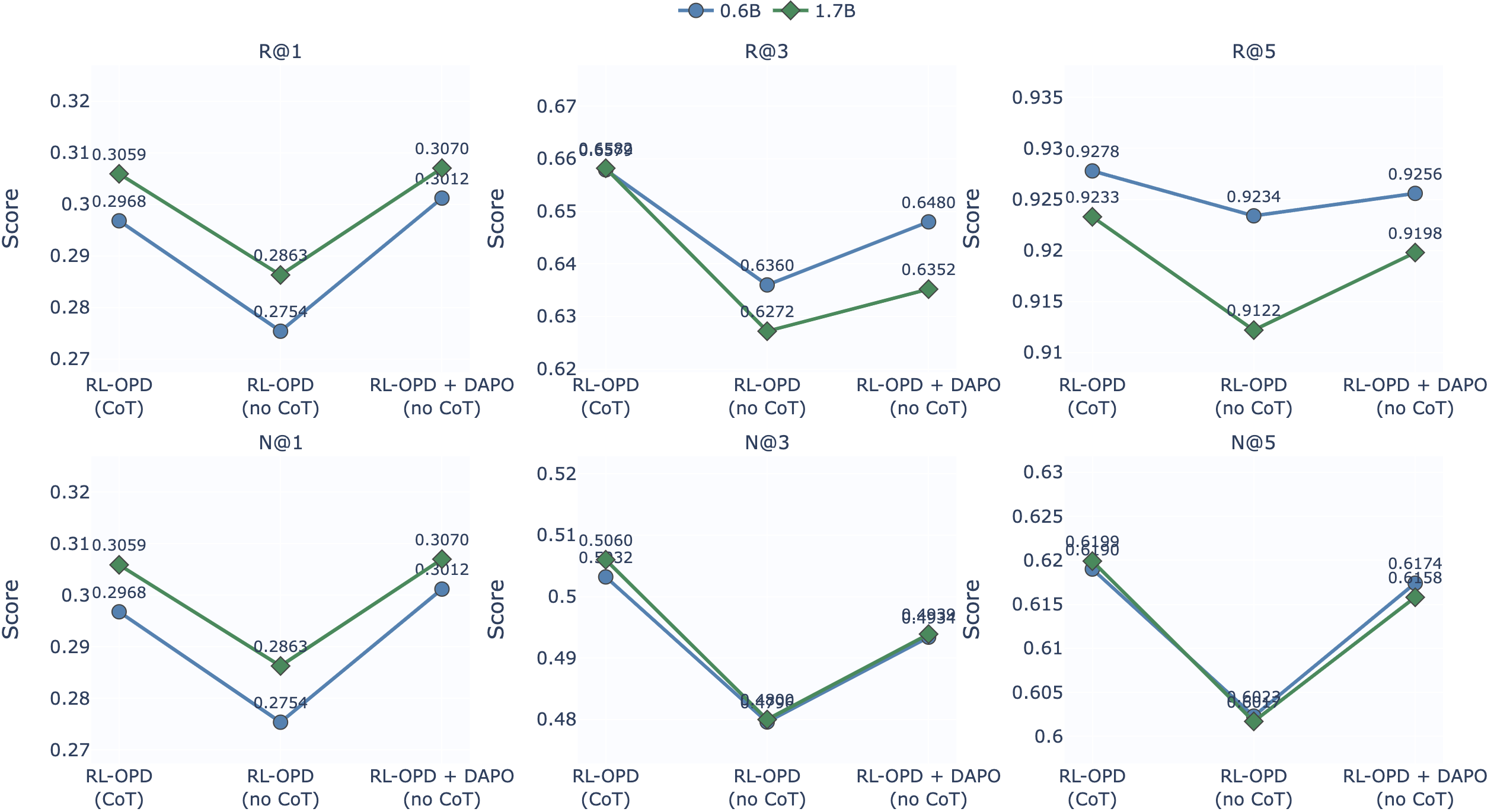}
    \caption{Recall and NDCG across RL-OPD with CoT (explicit reasoning traces), without CoT, and after a second RL pass that internalizes reasoning.}
    \label{fig:internal}
\end{figure}

\paragraph{Internalizing Reasoning (Fig.~\ref{fig:internal})}
To stress-test whether implicit reasoning can replace explicit CoT, we evaluate on a hard subset---sessions flagged by a large foundation LLM as requiring above-average reasoning, where bypassing CoT should hurt most. The first two points trace the \emph{same} RL-OPD checkpoint with and without CoT: removing the trace at inference drops Recall and NDCG sharply on both 0.6B and 1.7B students (\textbf{$-$6--7\%} R@1/N@1, \textbf{$-$3--5\%} R@3/N@3), confirming the CoT carries real ranking signal on this slice. A second RL pass closes the gap entirely and slightly surpasses the CoT reference at the head of the ranking (R@1/N@1: \textbf{0.3012} vs.\ \textbf{0.2968} for 0.6B; \textbf{0.3070} vs.\ \textbf{0.3059} for 1.7B), and matches it at deeper cutoffs (0.6B R@5: \textbf{0.9256} vs.\ \textbf{0.9278}; N@5: \textbf{0.6174} vs.\ \textbf{0.6190}, both within $\sim$0.3\%). The OPD-pretrained backbone already encodes a robust ranking prior over reasoning traces, so the second RL phase only needs to re-route this prior into the direct-output policy, internalizing CoT through the shared backbone. The deployed artifact is thus reasoning-free at iso- or super-CoT quality even on reasoning-heavy traffic.

\subsection{Reasoning Quality and Ablation}
\label{sec:reasoning_analysis}

\begin{figure}
    \centering
    \includegraphics[width=0.75\linewidth]{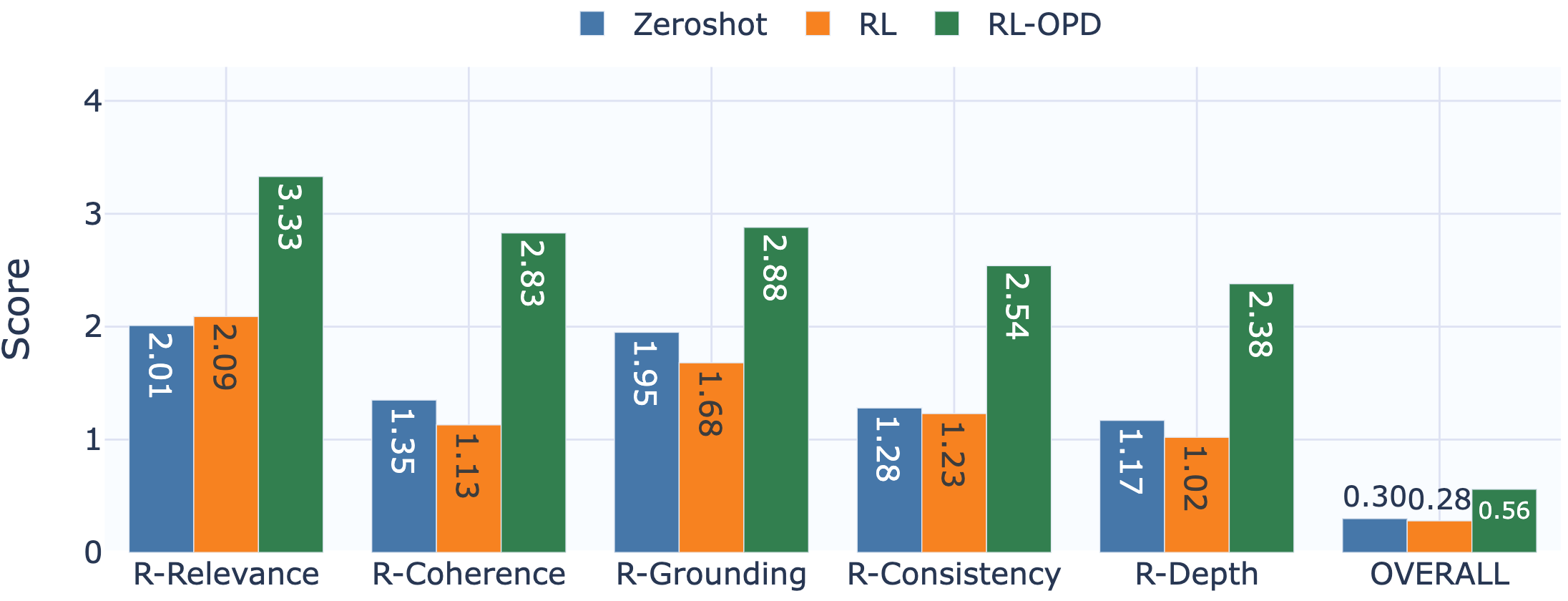}
    \caption{Reasoning quality (5 rubrics evaluated by LLM-as-a-judge using Claude CLI) of different GR2 variants.}
    \label{fig:reasoning_judge}
\end{figure}

\paragraph{Reasoning quality across training recipes (Fig.~\ref{fig:reasoning_judge}).}
Applying RL directly on top of the zero-shot checkpoint slightly degrades reasoning quality across all five rubrics relative to OPD-initialized variants; the degradation is sharpest on \textbf{Depth} (\mbox{$\approx$}$1.02$, near the rubric floor), confirming that RL alone learns to \emph{rank well without thinking well}.
The \textbf{RL-OPD} recipe recovers reasoning quality while preserving the ranking gains, producing traces that are simultaneously shorter and more focused.

\begin{figure}
    \centering
    \includegraphics[width=0.75\linewidth]{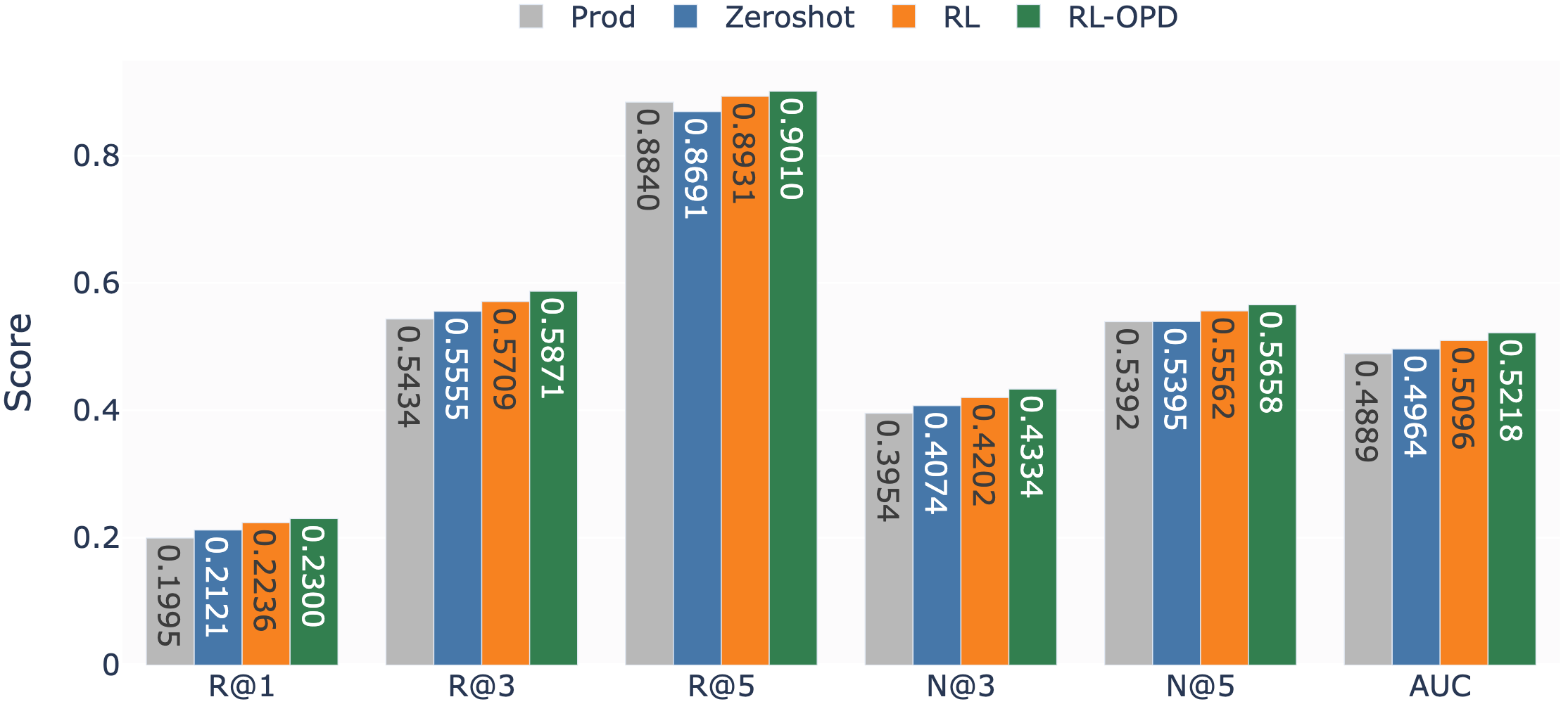}
    \caption{Ranking performance of different GR2 variants.}
    \label{fig:ablation}
\end{figure}

\begin{table*}[t]
  \centering
  \caption{A case study demonstrating the value of GR2 vs.\ the pointwise legacy baseline on a representative session. \textcolor{red}{Red} marks the clicked candidate. Special tokens and SID are removed for readability.}
  \label{tab:reasoning_case_study}
  \scriptsize
  \setlength{\tabcolsep}{6pt}
  \renewcommand{\arraystretch}{1.15}
  \begin{tabular}{@{}p{0.4\textwidth} p{0.22\textwidth} p{0.32\textwidth}@{}}
    \toprule
    \textbf{User interaction history} & \textbf{Legacy baseline (Top-6)} & \textbf{GR2 reasoning + Top-6 (reranked indices)} \\
    \midrule

    \begin{enumerate}[leftmargin=1.2em,nosep,label=\textbf{\arabic*.}]
      \item TRX AMBIDEXTROUS HOLSTER PRO (TRX Holster): The TRX HolsterTM Slim offers reliable compatibility and a snug fit for your Glock, P80, Smith \& Wess...
      \item Leather Drop Leg Holster with Light (Falco): Light-bearing Duty Drop Leg holster made of leather, specifically designed for professional use and ...
      \item Leather Drop Leg Holster with Light (Falco): Light-bearing Duty Drop Leg holster made of leather, specifically designed for professional use and ...
      \item Menseventwear Casual Men's Double Breasted Tweed Herringbone V Neck Waistcoat XS / Brown (Adam Reed): 20\% Wool / 80\% Polyester Herringbone V Neck 2 Real Welt pocket Double Breasted 8 Buttons Back metal ...
      \item US Flag \& Constitution Pistol Case (Relic Wood): US Flag \& Constitution Pistol Case The perfect pistol case for proud and patriotic handgun owners. I...
      \item Guides Choice® Speed Loader Pouch (Diamond D Custom Leather): Due To Popular Demand We Have Created What We Believe To Be A Very Ergonomic And Easy To Use Pouch F...
      \item Alaska Hunter Hip Holster (Diamond D Custom Leather): The Alaska Hunter Leather Gun Holster Is A Hip Holster With Ease Of Carry In Mind. We Have Placed Th...
      \item Leather Drop Leg Holster with Light (Falco): Light-bearing Duty Drop Leg holster made of leather, specifically designed for professional use and ...
      \item Wearable Recorder: Wearable Recorder
      \item John Wayne Cowboy Leather Vest - Brown Real Sheep Leather Waistcoat For Mens (Handmade)
      \item Minolta MND20 Point \& Shoot Digital Camera with SD Card (Minolta): Whether you're sightseeing, celebrating with family, or gathering with friends, this point and shoot...
      \item Gordon® 6 in. Bowie Knife (Gordon®): The GORDONTM 6 in. Bowie Knife is perfect for piercing and cutting in tight spaces. Stainless steel b...
      \item Jos. A. Bank Solid Watch Cap (Jos. A. Bank): A versatile essential for cooler weather, this watch cap is designed for both warmth and timeless st...
      \item HyperIce Hyperboot by Nike × (HyperIce): These wearable technology shoes offer three levels of heat and Normatec dynamic air compression to s...
      \item Alleviate Loft 2 Left Foot Brace (Alleviate): Get immediate relief from plantar fasciitis or posterior tibial tendonitis (PTTD) pain with the Alle...
    \end{enumerate}

     &

    \begin{enumerate}[leftmargin=1.2em,nosep,label=\textbf{\arabic*.}]
      \item Diesel Men9.5us Leather Lace-up Boots Dark Brown Distressed Finish Almost Unused (Diesel)
      \item Oakley Probation Polished Chrome No Lens Oo4041-02 Inmate Metal Rare (Oakley)
      \item 1-1/2" Domed Face German-style Chasing Hammer 11" Length Jewelry Making Tool (2)
      \item \textcolor{red}{70s-80ś Germany Police German Policeman Genuine Leather Leather Coat Vintage (Police)}
      \item Mens Leather Jeans Pants Trouser 5 Pockets Cowhide Brown Breeches Bluf Levis 501 (Dexa Apperals)
      \item Cnc Stainless Steel Replacement Safety Razor Stubby Short 3.0" Inch Handle Only (Unspecified)
    \end{enumerate}
    &
    <think>
    
    The user has a history of clicking on various products, mainly related to holsters, leather items, and some accessories. Let's look at their previous interactions to see what they like.
    They have a strong interest in leather holsters and cases (TRX, Falco, Diamond D), which are all related to firearms accessories. They also clicked on leather coats, like the John Wayne Vest and Menseventwear. So products related to leather and accessories would be a good fit.
    
    Looking at the products to predict:  
    \begin{enumerate}[leftmargin=1.2em,nosep,label=\textbf{\arabic*.}]
        \item The first one is Leather boots.
        \item Oakley is a brand they haven't clicked before, but maybe not relevant.
        \item The third one is a jewelry making tool. The user didn't click anything like this before.
        \item The forth one is similar to previous leather coats (John Wayne Cowboy Vest, Menseventwear Waistcoat) they clicked on.
        \item Mens Leather Jeans Pants – They clicked on leather pants/jeans in the past (like the Menseventwear Waistcoat).
        \item Cnc Stainless Steel Razor – The user didn't click on any razors before.
    \end{enumerate}
    
    Product 4 (Leather Coat) is similar in style. Product 5 (Leather Pants) is also a type of clothing. Product 1 (Leather Boots) might be a bit less relevant than pants but could be clicked if they like the style. Product 2 (Oakley Probation) is not leather and not firearms. Product 3 (Jewelry tool) and product 6 (Stainless Steel Razor) are not aligned with their past behavior.
    
    So the most likely clicked would be 4 and 5, followed by 1. Product 2 (Oakley), 3 (tool), and 6 (razor) might be lower on the list. 
    
    </think>
    
    [4, 5, 1, 2, 3, 6]
 
    \\
    \bottomrule
  \end{tabular}
\end{table*}

\paragraph{Ablation: training recipe trade-offs (Fig.~\ref{fig:ablation}).}
\label{sec:ablation}
We compare three recipes side-by-side on both ranking metrics and reasoning quality:
(i) RL-only on the zero-shot checkpoint, and
(ii) \textbf{RL-OPD}.
RL-only matches \textbf{RL-OPD} on ranking metrics but underperforms on every reasoning rubric;
\textbf{RL-OPD} Pareto-dominates both, supporting our headline claim that OPD supplies the reasoning prior while RL sharpens the ranking objective on top of it.

\paragraph{Case study (Table~\ref{tab:reasoning_case_study}).}
GR2 grounds its ranking in product attributes (category, material, function) and the user's revealed click patterns---semantics the sparse-ID baseline cannot represent. From a history dominated by leather holsters, pistol cases, and leather outerwear, GR2 infers a coherent ``leather goods + firearms accessories'' profile and promotes the vintage leather police coat to the top, recovering the click that the baseline ranks fourth. Non-leather, off-theme candidates (jewelry-making hammer, safety razor, Oakley sunglasses) are correctly pushed to the bottom, while the leather boots and leather pants are kept in the middle as plausible but weaker matches.

\section{Related Works}\label{sec:related}

\paragraph{Generative LLM RecSys}
Recent work increasingly reformulates recommendation as sequence generation. TIGER~\citep{rajput2023recommender} pioneered the use of Semantic IDs (SIDs), discretizing item representations via RQ-VAE to enable autoregressive recommendation without large embedding tables. OneRec~\citep{zhou2025onerec} cast multiple recommendation tasks into a unified text-to-text LLM framework, and OneRec-Think~\citep{liu2025onerec} extended it with explicit reasoning traces, showing that chain-of-thought intermediate steps improve accuracy and robustness. OpenOneRec~\citep{zhou2025openonerec} contributes a reproducible research framework for this family. PLUM~\citep{he2025plum} similarly uses continued pre-training to align LLMs with the recommendation domain and adds a task-specific fine-tuning objective for generative retrieval.
GR2 inherits the SID and item-alignment ideas but differs in three respects: (1) a tokenizer that achieves substantially higher SID uniqueness, (2) high-quality reasoning traces produced via targeted prompting and rejection sampling, and (3) a DAPO-based reward optimization.

\paragraph{Reasoning LLM}
Chain-of-thought (CoT) prompting elicits multi-step reasoning by generating intermediate steps before final answers~\citep{wei2022chain}, and subsequent work has formalized and extended reasoning structures beyond CoT~\citep{xia2025beyond,wang2024strategic}. RL methods such as GRPO and its variants further improve reasoning faithfulness and coherence via verifiable rewards~\citep{shao2024deepseekmath,yu2025dapo,zheng2025group,chen2025minimax}, providing the broader context for our optimization strategy.
A close work is LLM-based document re-ranking. ReaRank reasons before re-ranking passage lists via RL~\citep{zhang2025rearank}; Rank-R1 applies RL-based reasoning optimization to document ranking~\citep{feng2025iranker}; ReasonRank and MM-R5 extend reasoning-augmented reranking to multi-view and multi-modal settings~\citep{xu2025mm}; $R^4ec$~\citep{gu2025r} adds iterative reflection and refinement; and \citet{wang2025llm} use a candidate-randomization bootstrap to improve fairness in LLM-based reranking. GR2 differs in its focus on semantic IDs and reward functions tailored to recommendation re-ranking, combined with structured reasoning traces and rejection sampling for higher-quality supervision.

\section{Conclusion}\label{sec:con}
We presented \textbf{GR2}, a Generative Reasoning Re-Ranker that brings LLM world knowledge and chain-of-thought reasoning to the final re-ranking stage of large-scale recommendation systems. GR2 combines mid-training on SIDs, reasoning-trace distilled from a stronger teacher, and RL with verifiable rewards. To make the recipe industry-viable, we further introduced a context compressor that cuts input length by $>$80\% at iso-quality, OPD as a scalable alternative to SFT, and reasoning-internalizing CoT distillation that delivers a $\sim$15$\times$ serving-ROI gain.
Across industrial-scale traffic, GR2 improves over internal baselines (+18.7\% R@1, +7.1\% R@3; +9.6\% N@3), with gains robust to test-set scale, two-week traffic staleness, and increasing model size. Ablations and LLM-as-judge analysis confirm that OPD supplies the reasoning prior while RL sharpens the ranking objective on top of it---neither is sufficient alone.

\clearpage
\newpage
\bibliographystyle{assets/plainnat}
\bibliography{paper}



\end{document}